\documentclass[a4paper,11pt]{article}
\usepackage{pos}

\def\beq{\begin{equation}}
\def\eeq{\end{equation}}
\def\beqa{\begin{eqnarray}}
\def\eeqa{\end{eqnarray}}

\title{Higher-order corrections for $t{\bar t}Z$ production}

\author*{Nikolaos Kidonakis}
\author{Chris Foster}

\affiliation{Department of Physics, Kennesaw State University,\\
Kennesaw, GA 30144, USA}

\emailAdd{nkidonak@kennesaw.edu}

\abstract{We present calculations of higher-order QCD and electroweak (EW) corrections to the associated production of a top-antitop quark pair and a $Z$ boson, i.e. $t{\bar t}Z$ production, in proton collisions. We find that the contributions from soft-gluon corrections are numerically dominant and large. We present approximate NNLO (aNNLO) and approximate N$^3$LO (aN$^3$LO) cross sections that include soft corrections on top of the exact NLO QCD result. We also add electroweak corrections through NLO. We compare our aN$^3$LO QCD + NLO EW theoretical results to measurements of total cross sections from the LHC, and we also calculate top-quark differential distributions in transverse momentum and rapidity.}

\FullConference{XXXII International Workshop on Deep Inelastic Scattering and Related Subjects (DIS2025)\\
24-28 March, 2025\\
Cape Town, South Africa\\}

\begin{document}
\maketitle

\section{Introduction}

The production of a top-antitop pair with a $Z$ boson, i.e. $t{\bar t}Z$ production, is a process that is important for measuring the coupling of the top quark to the $Z$ boson and, thus, for the search for new physics. Total $t{\bar t}Z$ cross sections have been measured at the LHC at 7 TeV, 8 TeV, and 13 TeV \cite{CMS,ATLAS} energies. 

The NLO QCD corrections for this process were calculated in Refs. \cite{LMMP,KTP} and found to be large, while electroweak corrections were computed in \cite{FHPSZ}. Calculations at yet higher orders are needed for better theoretical predictions of the production cross sections. An important set of higher-order corrections arises from the emission of soft gluons near partonic threshold. We use the soft-gluon resummation formalism of Refs. \cite{NKGS,NKtt,NKloop} which was applied to top-antitop pair production \cite{NKtt,NKloop,NKttaN3LO} and single-top production \cite{NKsingletop}, and was extended \cite{FK2020} to $2 \to 3$ processes in single-particle-inclusive (1PI) kinematics, in particular $tqH$ production \cite{FK2021}, $tq\gamma$ production \cite{NKNY2022}, $tqZ$ production \cite{NKNYtqZ}, $t{\bar t} \gamma$ production \cite{NKAT}, $t{\bar t}W$ production \cite{KFttW}, $t{\bar t}Z$ production \cite{KFttZ}, and $t{\bar t}H$ production \cite{NKNYttH}. We find that, as for all the processes listed above, the soft-gluon corrections in $t{\bar t}Z$ production \cite{KFttZ} account for the majority of the complete corrections. Other resummation formalisms for $t{\bar t}Z$ production, using different kinematics, i.e. the invariant mass of $t{\bar t}Z$, were used in Refs. \cite{BFOPS,KMSST,BFFPPT}. 

By expanding the resummed cross section and matching to NLO, we calculate approximate NNLO (aNNLO) and approximate N$^3$LO (aN$^3$LO) cross sections and top-quark differential distributions for $t{\bar t}Z$ production. We further add electroweak (EW) corrections through NLO, thus making aN$^3$LO QCD + NLO EW predictions \cite{KFttZ}.

\section{Resummation of soft-gluon corrections for $t{\bar t}Z$ production}

We start with a discussion of soft-gluon resummation for $t{\bar t}Z$ production in 1PI kinematics, with the top quark being the observed particle. At leading order, the parton-level processes are $a(p_a)+b(p_b) \to t(p_t)+{\bar t}(p_{\bar t})+Z(p_Z)$, where $a$ and $b$ denote the incoming partons (quarks and antiquarks or gluons), and we define $s=(p_a+p_b)^2$, $t=(p_a-p_t)^2$, and $u=(p_b-p_t)^2$, as well as $s'=(p_t+p_{\bar t})^2$, $t'=(p_b-p_{\bar t})^2$, and $u'=(p_a-p_{\bar t})^2$. With $m_t$ the top-quark mass and $p_g$ the momentum of an additional emitted gluon, we define a threshold variable $s_4=(p_{\bar t}+p_Z+p_g)^2-(p_{\bar t}+p_Z)^2=s+t+u-m_t^2-(p_{\bar t}+p_Z)^2$, which vanishes as $p_g \to 0$. The soft-gluon corrections appear in terms involving ``plus distributions''  of $\ln^k(s_4/m_t^2)/s_4$ with $k$ an integer from 0 to $2n-1$ for the $n$th order corrections.

Soft-gluon resummation follows from the factorization properties of the cross section and the renormalization-group evolution (RGE) of functions that describe soft and collinear emission \cite{NKGS,NKtt,FK2020,GS,LOS}. We write the differential cross section for $t{\bar t}Z$ production as 
\beq
d\sigma_{pp \to t{\bar t}Z}=\sum_{a,b} \; 
\int dx_a \, dx_b \,  \phi_{a/p}(x_a, \mu_F) \, \phi_{b/p}(x_b, \mu_F) \, 
d{\hat \sigma}_{ab \to t{\bar t}Z}(s_4, \mu_F) \, ,
\label{sigmattZ}
\eeq
where $\phi_{a/p}$  and $\phi_{b/p}$ are parton distribution functions (pdf) in the proton, ${\hat \sigma}_{ab \to t{\bar t}Z}$ is the partonic cross section, and $\mu_F$ is the factorization scale.

The cross section factorizes under Laplace transforms, which are given by
\beq
 d{\tilde{\hat\sigma}}_{ab \to t{\bar t}Z}(N)=\int_0^{s_4^{\rm max}} \frac{ds_4}{s} \,  e^{-N s_4/s} \; d{\hat\sigma}_{ab \to t{\bar t}Z}(s_4), 
\eeq
with $N$ the transform variable. Under Laplace transforms, logarithms of $s_4$ turn into logarithms of $N$ which exponentiate. Replacing the colliding protons by partons in Eq. (\ref{sigmattZ}) \cite{NKGS,FK2020,GS}, and defining ${\tilde \phi}(N)=\int_0^1 e^{-N(1-x)} \phi(x) \, dx$, we find a factorized form
\beq
d{\tilde \sigma}_{ab \to t{\bar t}Z}(N)= {\tilde \phi}_{a/a}(N_a, \mu_F) \, {\tilde \phi}_{b/b}(N_b, \mu_F) \, d{\tilde{\hat \sigma}}_{ab \to t{\bar t}Z}(N, \mu_F) \, .
\label{factor}
\eeq

The cross section is further refactorized \cite{NKGS,NKtt,FK2020} via an infrared-safe short-distance hard function, $H_{ab \to t{\bar t}Z}$, and a soft function, $S_{ab \to t{\bar t}Z}$, which describes the emission of noncollinear soft gluons. Both $H_{ab \to t{\bar t}Z}$ and $S_{ab \to t{\bar t}Z}$ are matrices in color space. We have
\beq
d{\tilde{\sigma}}_{ab \to t{\bar t}Z}(N)={\tilde \psi}_{a/a}(N_a,\mu_F) \, {\tilde \psi}_{b/b}(N_b,\mu_F) \, {\rm tr} \left\{H_{ab \to t{\bar t}Z} \left(\alpha_s(\mu_R)\right) \, {\tilde S}_{ab \to t{\bar t}Z} \left(\frac{\sqrt{s}}{N \mu_F} \right)\right\} \, ,
\label{refactor}
\eeq
where the functions $\psi$ are distributions for incoming partons at fixed momentum \cite{NKGS,GS} and $\mu_R$ is the renormalization scale. Comparing Eqs. (\ref{factor}) and (\ref{refactor}), we find 
\beq
d{\tilde{\hat \sigma}}_{ab \to t{\bar t}Z}(N)=
\frac{{\tilde \psi}_{a/a}(N_a, \mu_F) \, {\tilde \psi}_{b/b}(N_b, \mu_F)}{{\tilde \phi}_{a/a}(N_a, \mu_F) \, {\tilde \phi}_{b/b}(N_b, \mu_F)} \; \,  {\rm tr} \left\{H_{ab \to t{\bar t}Z}\left(\alpha_s(\mu_R)\right) \, {\tilde S}_{ab \to t{\bar t}Z}\left(\frac{\sqrt{s}}{N \mu_F} \right)\right\} \, .
\label{sigNttZ}
\eeq

The soft function, ${\tilde S}_{ab \to t{\bar t}Z}$, obeys a renormalization-group equation with a soft anomalous dimension matrix, $\Gamma_{\! S \, ab \to t{\bar t}Z}$, which is calculated from the coefficients of the ultraviolet poles of the relevant eikonal diagrams \cite{NKGS,NKtt,NKloop,NKsingletop,FK2020}.

The $N$-space resummed cross section is derived from the RGE of ${\tilde S}_{ab \to t{\bar t}Z}$, ${\tilde \psi}_{a/a}$, ${\tilde \psi}_{b/b}$, ${\tilde \phi}_{a/a}$, and ${\tilde \phi}_{b/b}$ in Eq. (\ref{sigNttZ}), and it is given by
\beqa
d{\tilde{\hat \sigma}}_{ab \to t{\bar t}Z}^{\rm resum}(N) &=&
\exp\left[\sum_{i=a,b} E_{i}(N_i)\right] \, 
\exp\left[\sum_{i=a,b} 2 \int_{\mu_F}^{\sqrt{s}} \frac{d\mu}{\mu} \gamma_{i/i}(N_i)\right]
\nonumber\\ && \hspace{-15mm} \times \,
{\rm tr} \left\{H_{ab \to t{\bar t}Z}\left(\alpha_s(\sqrt{s})\right) {\bar P} \exp \left[\int_{\sqrt{s}}^{{\sqrt{s}}/N}
\frac{d\mu}{\mu} \; \Gamma_{\! S \, ab \to t{\bar t}Z}^{\dagger} \left(\alpha_s(\mu)\right)\right] \; \right.
\nonumber\\ && \hspace{-5mm} \left. \times \,
{\tilde S}_{ab \to t{\bar t}Z} \left(\alpha_s\left(\frac{\sqrt{s}}{N}\right)\right) \;
P \exp \left[\int_{\sqrt{s}}^{{\sqrt{s}}/N}
\frac{d\mu}{\mu}\; \Gamma_{\! S \, ab \to t{\bar t}Z}
\left(\alpha_s(\mu)\right)\right] \right\} \, .
\label{resummedcs}
\eeqa
The first exponential resums soft and collinear gluon emission from the incoming partons \cite{GS}, the second involves the parton anomalous dimensions, followed by exponentials with $\Gamma_{\! S \, ab \to t{\bar t}Z}$. 

The soft anomalous dimensions for $t{\bar t} Z$ production via quark/antiquark or gluon processes are basically the same as for $t{\bar t}$ production \cite{NKGS,NKtt,NKloop} because the color structure of the hard scattering is the same, with only minor modifications in consideration of the $2 \to 3$ kinematics. For the $g(p_a)+g(p_b) \to t(p_t)+{\bar t}(p_{\bar t})+Z(p_Z)$ channel, in the color tensor basis $c_1^{gg\rightarrow t{\bar t}Z}=\delta^{ab}\,\delta_{12}$, $c_2^{gg\rightarrow t{\bar t}Z}=d^{abc}\,T^c_{12}$, $c_3^{gg\rightarrow t{\bar t}Z}=i f^{abc}\,T^c_{12}$, the elements of the $3\times 3$ matrix $\Gamma_{\! S \, gg\rightarrow t{\bar t}Z}$ are given at one loop by
\beqa
&& \hspace{-5mm} \Gamma_{11}^{(1)}= \Gamma_{\rm cusp}^{(1)}  \, , \quad \Gamma_{12}^{(1)}=\Gamma_{21}^{(1)}=0 \, , \quad \Gamma_{13}^{(1)}= \frac{1}{2}\ln\left(\frac{(t-m_t^2)(t'-m_t^2)}{(u-m_t^2)(u'-m_t^2)}\right) \, , \quad \Gamma_{31}^{(1)} = 2 \, \Gamma_{13}^{(1)} \, ,
\nonumber \\ && \hspace{-5mm}
\Gamma_{22}^{(1)}= \left(1-\frac{C_A}{2C_F}\right) \Gamma_{\rm cusp}^{(1)}
+\frac{C_A}{2}\left[\frac{1}{2}\ln\left(\frac{(t-m_t^2)(t'-m_t^2)(u-m_t^2)(u'-m_t^2)}{s^2\, m_t^4}\right)-1\right] \, ,
\nonumber \\ && \hspace{-5mm}
\Gamma_{23}^{(1)}=\frac{C_A}{2} \, \Gamma_{13}^{(1)} \, , \quad 
\Gamma_{32}^{(1)}=\frac{(N_c^2-4)}{2N_c} \, \Gamma_{13}^{(1)} \, , \quad
\Gamma_{33}^{(1)}=\Gamma_{22}^{(1)} \, ,
\eeqa
where $\Gamma_{\rm cusp}^{(1)}$ is the one-loop QCD massive cusp anomalous dimension. At two loops, 
\beqa
&& \hspace{-5mm} \Gamma_{11}^{(2)}= \Gamma_{\rm cusp}^{(2)} \, , \quad \Gamma_{12}^{(2)}=\Gamma_{21}^{(2)}=0 \, , \quad \Gamma_{13}^{(2)}=\left(K_2-C_A N_2\right) \Gamma_{13}^{(1)} \, , \quad 
\Gamma_{31}^{(2)}=\left(K_2+C_A N_2\right) \Gamma_{31}^{(1)} \, ,
\nonumber \\ && \hspace{-5mm}
\Gamma_{22}^{(2)}= K_2 \, \Gamma_{22}^{(1)}
+\left(1-\frac{C_A}{2C_F}\right) \left(\Gamma_{\rm cusp}^{(2)}-K_2 \Gamma_{\rm cusp}^{(1)}\right)+\frac{C_A^2}{4}(1-\zeta_3) \, ,
\nonumber \\ && \hspace{-5mm}
\Gamma_{23}^{(2)}= K_2 \, \Gamma_{23}^{(1)} \, , \quad
\Gamma_{32}^{(2)}= K_2 \, \Gamma_{32}^{(1)} \, , \quad
\Gamma_{33}^{(2)}=\Gamma_{22}^{(2)} \, , 
\eeqa
where $\Gamma_{\rm cusp}^{(2)}$ is the two-loop QCD massive cusp anomalous dimension \cite{NKloop}. Expressions for $K_2$, $N_2$, and $\Gamma_{\rm cusp}^{(2)}$ in the above equation, as well as corresponding expressions for the soft anomalous dimension matrix for the quark-antiquark channel, can be found in our related paper on $t{\bar t}W$ production \cite{KFttW}.

We expand the resummed cross section, Eq. (\ref{resummedcs}), to N$^3$LO, and then invert back to momentum space. We add the second-order soft-gluon corrections to the exact NLO results to calculate aNNLO predictions, and we further add the third-order soft-gluon corrections to make aN$^3$LO predictions.

\section{Total and differential cross sections}

In this section, we present numerical results for the $t{\bar t}Z$ production total cross section as well as the top-quark transverse momentum ($p_T$) and rapidity distributions. We set the factorization and renormalization scales equal to each other, and use as central scale $\mu=m_t=172.5$ GeV. The complete NLO QCD and NLO QCD + NLO EW results are calculated with {\small \sc MadGraph5\_aMC@NLO} \cite{MG5,MGew}. We derive approximate results at NNLO and N$^3$LO by adding second-order and third-order soft-gluon corrections to NLO. We use MSHT20 NNLO pdf \cite{MSHT20NNLO} for all results presented here.

\begin{figure}[htbp]
\begin{center}
\includegraphics[width=62mm]{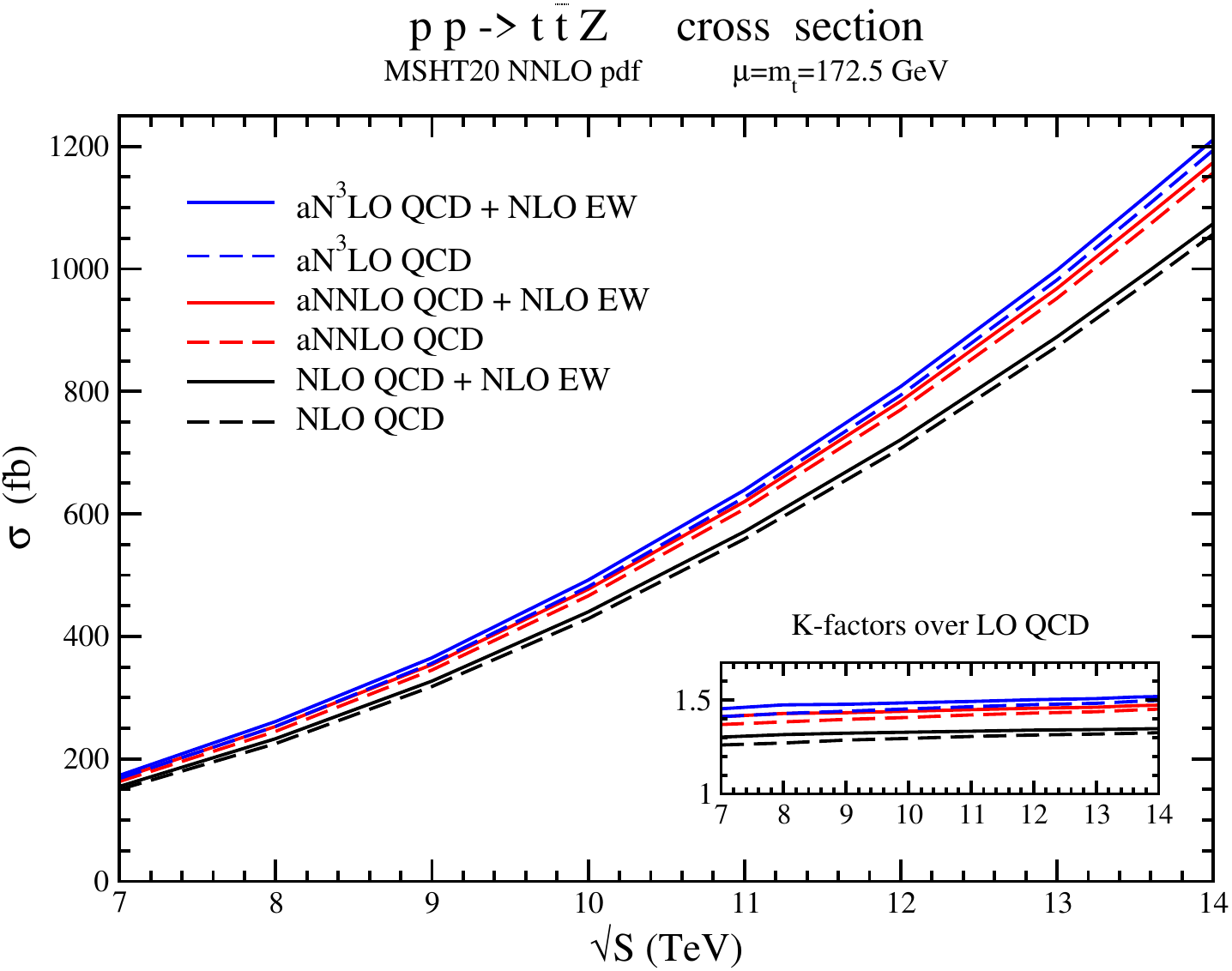}
\hspace{10mm}
\includegraphics[width=62mm]{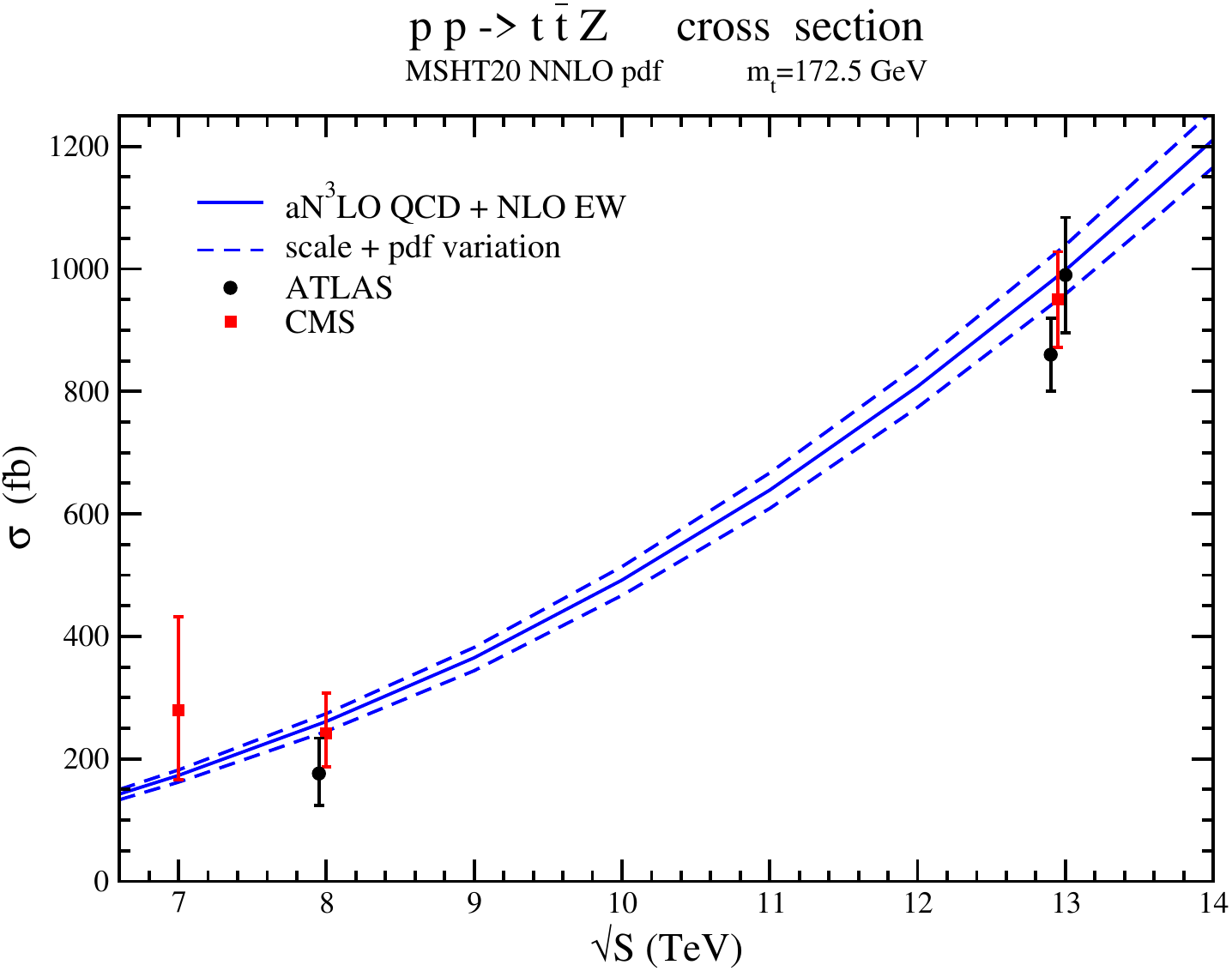}
\caption{The total cross sections for $t{\bar t}Z$ production in $pp$ collisions. The left plot shows cross sections at orders from NLO QCD through aN$^3$LO QCD + NLO EW, and the right plot shows a comparison with LHC data at 7, 8, and 13 TeV energies \cite{CMS,ATLAS}.}
\label{ttZprcs}
\end{center}
\end{figure}

In Fig. \ref{ttZprcs}, we show our predictions for the $t{\bar t}Z$ total cross section in $pp$ collisions for the LHC energy range. The plot on the left shows results at various perturbative orders from NLO QCD through aN$^3$LO QCD + NLO EW. The inset plot on the left shows the $K$-factors, i.e. the ratios of the higher-orders predictions relative to LO QCD. The aN$^3$LO QCD + NLO EW $K$-factor is 1.51 at 13 TeV and 1.52 at 13.6 TeV. The plot on the right compares the aN$^3$LO QCD + NLO EW prediction, including scale and pdf uncertainties, with data from ATLAS \cite{ATLAS} and CMS \cite{CMS}.

\begin{figure}[htbp]
\begin{center}
\includegraphics[width=62mm]{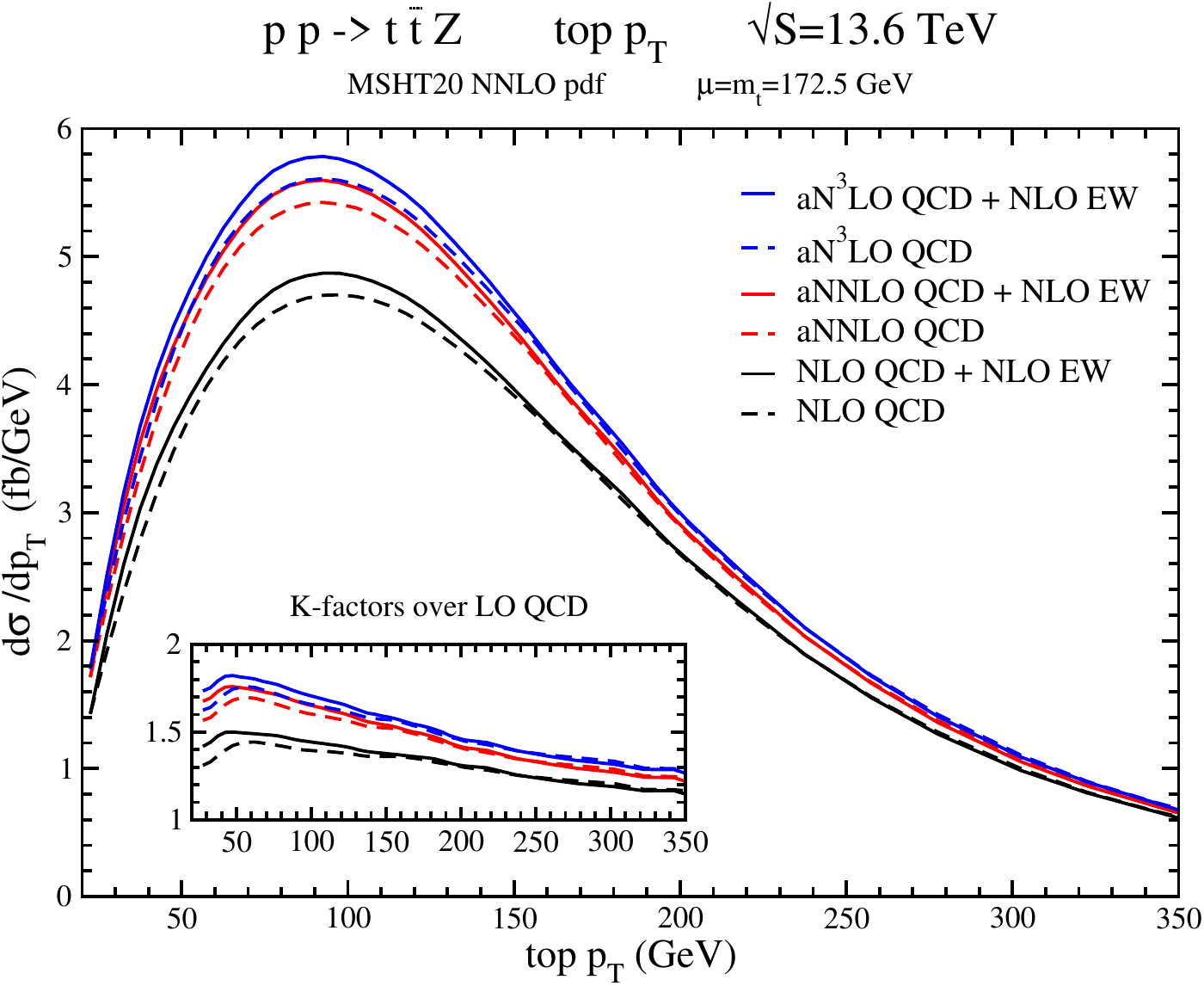}
\hspace{10mm}
\includegraphics[width=62mm]{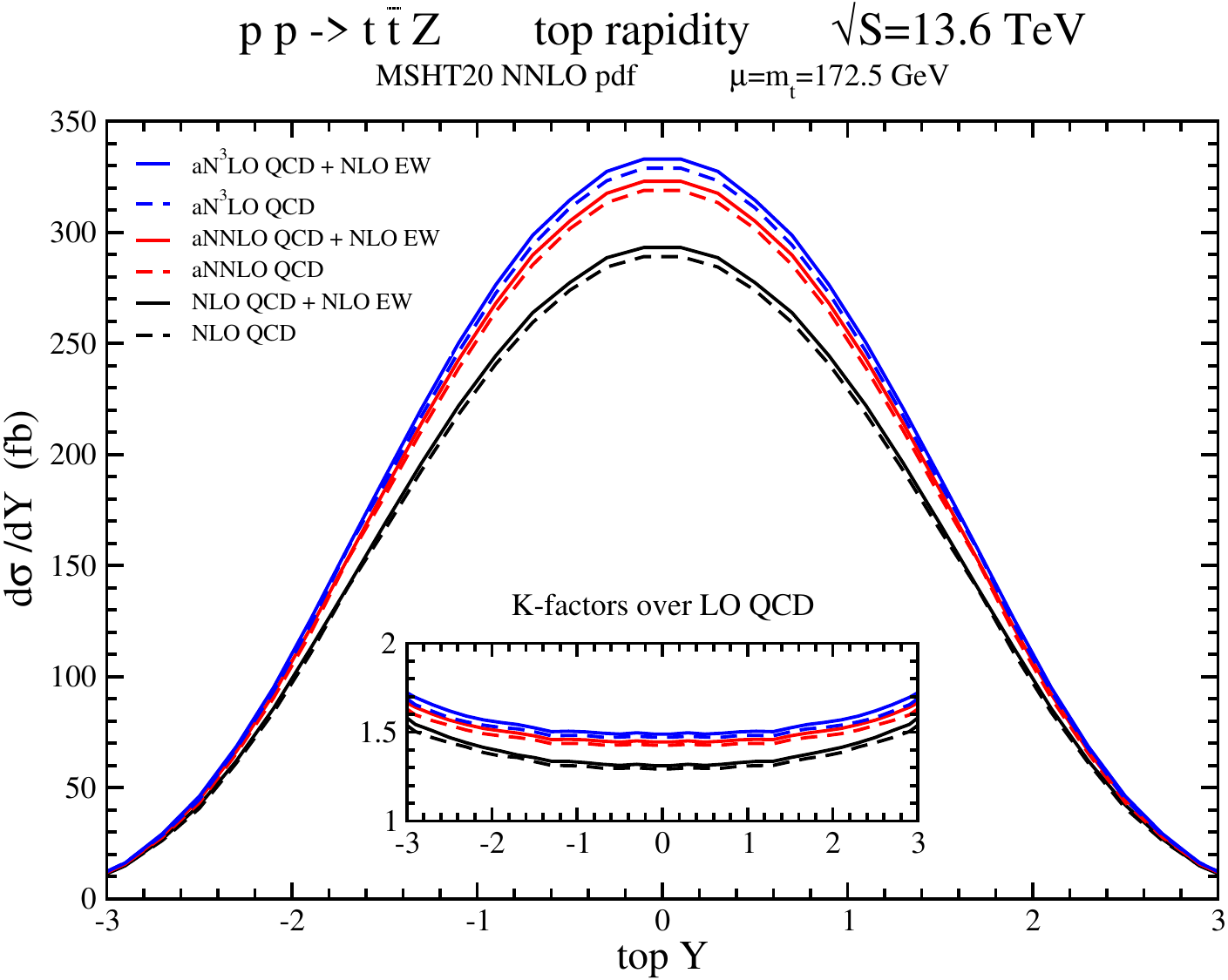}
\caption{The top-quark $p_T$ (left) and rapidity (right) distributions at orders from NLO QCD through aN$^3$LO QCD + NLO EW for $t{\bar t}Z$ production in $pp$ collisions at LHC energy of 13.6 TeV.}
\label{ptytop}
\end{center}
\end{figure}

In Fig. \ref{ptytop}, we display the top-quark $p_T$ (left plot) and rapidity (right plot) distributions for $t{\bar t} Z$ production in $pp$ collisions at 13.6 TeV LHC energy. Results are shown from NLO QCD through aN$^3$LO QCD + NLO EW with $\mu=m_t$. The inset plots show the $K$-factors of the higher-order results relative to LO QCD. The higher-order contributions are very significant, giving as high as around 80\% enhancement over LO QCD for the $p_T$ and rapidity ranges shown in the plots.

\acknowledgments

This material is based upon work supported by the National Science Foundation under Grant No. PHY 2412071.


\begin{thebibliography}{99}

\bibitem{CMS}
CMS Collaboration, Phys. Rev. Lett. {\bf 110}, 172002 (2013) [arXiv:1303.3239]; Eur. Phys. J. C {\bf 74}, 3060 (2014) [arXiv:1406.7830]; JHEP {\bf 01}, 096 (2016) [arXiv:1510.01131]; JHEP {\bf 08}, 011 (2018) [arXiv:1711.02547]; JHEP {\bf 03}, 056 (2020) [arXiv:1907.11270]; JHEP {\bf 12}, 083 (2021) [arXiv:2107.13896]; Phys. Rev. D {\bf 108}, 032008 (2023) [arXiv:2208.12837].

\bibitem{ATLAS}
ATLAS Collaboration, JHEP {\bf 11}, 172 (2015) [arXiv:1509.05276]; Eur. Phys. J. C {\bf 77}, 40 (2017) [arXiv:1609.01599]; Phys. Rev. D {\bf 99}, 072009 (2019) [arXiv:1901.03584]; Eur. Phys. J. C {\bf 81}, 737 (2021) [arXiv:2103.12603]; JHEP {\bf 07}, 163 (2024) [arXiv:2312.04450].

\bibitem{LMMP}
A. Lazopoulos {\it et al.}, Phys. Lett. B {\bf 666}, 62 (2008) [arXiv:0804.2220].

\bibitem{KTP}
A. Kardos, Z. Trocsanyi, and C.G. Papadopoulos, Phys. Rev. D {\bf 85}, 054015 (2012) [arXiv:1111.0610].

\bibitem{FHPSZ}
S. Frixione {\it et al.}, JHEP {\bf 06}, 184 (2015) [arXiv:1504.03446].

\bibitem{NKGS}
N. Kidonakis and G. Sterman, Phys. Lett. B {\bf 387}, 867 (1996); Nucl. Phys. B {\bf 505}, 321 (1997) [arXiv:hep-ph/9705234].

\bibitem{NKtt}
N. Kidonakis, Phys. Rev. D {\bf 64}, 014009 (2001) [arXiv:hep-ph/0010002]; Mod. Phys. Lett. A {\bf 19}, 405 (2004) [arXiv:hep-ph/0401147]; Phys. Rev. D {\bf 73}, 034001 (2006) [arXiv:hep-ph/0509079].    

\bibitem{NKloop}
N. Kidonakis, Phys. Rev. Lett. {\bf 102}, 232003 (2009) [arXiv:0903.2561]; Phys. Rev. D {\bf 82}, 114030 (2010) [arXiv:1009.4935]; Phys. Rev. D {\bf 107}, 054006 (2023) [arXiv:2301.05972]. 

\bibitem{NKttaN3LO}
N. Kidonakis, Phys. Rev. D {\bf 90}, 014006 (2014) [arXiv:1405.7046]; Phys. Rev. D {\bf 91}, 031501 (2015) [arXiv:1411.2633]; Phys. Rev. D {\bf 101}, 074006 (2020) [arXiv:1912.10362]. 

\bibitem{NKsingletop}
N. Kidonakis, Phys. Rev. D {\bf 74}, 114012 (2006) [arXiv:hep-ph/0609287]; Phys. Rev. D {\bf 75}, 071501 (2007) [arXiv:hep-ph/0701080]; Phys. Rev. D {\bf 81}, 054028 (2010) [arXiv:1001.5034]; Phys. Rev. D {\bf 82}, 054018 (2010) [arXiv:1005.4451]; Phys. Rev. D {\bf 83}, 091503 (2011) [arXiv:1103.2792]; Phys. Rev. D {\bf 96}, 034014 (2017) [arXiv:1612.06426]; Phys. Rev. D {\bf 99}, 074024 (2019) [arXiv:1901.09928]; arXiv:2502.17631 (2025). 
 
\bibitem{FK2020}
M. Forslund and N. Kidonakis, Phys. Rev. D {\bf 102}, 034006 (2020) [arXiv:2003.09021].

\bibitem{FK2021}
M. Forslund and N. Kidonakis, Phys. Rev. D {\bf 104}, 034024 (2021) [arXiv:2103.01228].

\bibitem{NKNY2022}
N. Kidonakis and N. Yamanaka, Eur. Phys. J. C {\bf 82}, 670 (2022) [arXiv:2201.12877].

\bibitem{NKNYtqZ}
N. Kidonakis and N. Yamanaka, Phys. Lett. B {\bf 838}, 137708 (2023) [arXiv:2210.09542].

\bibitem{NKAT}
N. Kidonakis and A. Tonero, Phys. Rev. D {\bf 107}, 034013 (2023) [arXiv:2212.00096].

\bibitem{KFttW}
N. Kidonakis and C. Foster, Phys. Lett. B {\bf 854}, 138708 (2024) [arXiv:2312.00861]; N. Kidonakis, PoS(ICHEP2024)371 [arXiv:2408.13870]. 

\bibitem{KFttZ}
N. Kidonakis and C. Foster, Phys. Lett. B {\bf 860}, 139146 (2025) [arXiv:2410.01214].

\bibitem{NKNYttH}
N. Kidonakis and N. Yamanaka, arXiv:2509.23293.

\bibitem{BFOPS}
A. Broggio {\it et al.}, JHEP {\bf 04}, 105 (2017) [arXiv:1702.00800].

\bibitem{KMSST}
A. Kulesza {\it et al.}, Eur. Phys. J. C {\bf 79}, 249 (2019) [arXiv:1812.08622].

\bibitem{BFFPPT}
A. Broggio {\it et al.}, JHEP {\bf 08}, 039 (2019) [arXiv:1907.04343].

\bibitem{GS}
G. Sterman, Nucl. Phys. B {\bf 281}, 310 (1987). 

\bibitem{LOS}
E. Laenen, G. Oderda, and G. Sterman, Phys. Lett. B {\bf 438}, 173 (1998) [arXiv:hep-ph/9806467].

\bibitem{MG5}
J. Alwall {\it et al.}, JHEP {\bf 07}, 079 (2014) [arXiv:1405.0301].

\bibitem{MGew}
R. Frederix {\it et al.}, JHEP {\bf 07}, 185 (2018) [arXiv:1804.10017].

\bibitem{MSHT20NNLO}
S. Bailey {\it et al.}, Eur. Phys. J. C {\bf 81}, 341 (2021) [arXiv:2012.04684].

\end{thebibliography}
\end{document}